\documentclass[conference]{IEEEtran}
\usepackage{amsmath}
\usepackage{graphicx,color}
\usepackage{amsmath,amssymb}
\usepackage{bm}
\usepackage{url}
\usepackage{ascmac}

\newtheorem{theorem}{Theorem}

\IEEEsettopmargin{t}{0.8in}

\usepackage{comment}

\title{Chebyshev Inertial Landweber Algorithm \\ for Linear Inverse Problems}

\author{%
  \IEEEauthorblockN{
  		Tadashi Wadayama
  		and Satoshi Takabe}
  \IEEEauthorblockA{\IEEEauthorrefmark{1}%
		Nagoya Institute of Technology,
		Gokiso, Nagoya, Aichi, 466-8555, Japan,\\
 		\{wadayama, s\_takabe\}@nitech.ac.jp}
}

\begin{document}
%
\maketitle

\begin{abstract}
The Landweber algorithm defined on complex/real Hilbert spaces is a gradient descent algorithm for linear inverse 
problems. Our contribution is to present a novel method for accelerating convergence of the Landweber algorithm.
In this paper, we first extend the theory of the Chebyshev inertial iteration to the Landweber algorithm on Hilbert spaces. 
An upper bound on the convergence rate clarifies the speed of global convergence of the proposed method.
The Chebyshev inertial Landweber algorithm can be applied to wide class of signal recovery
problems on a Hilbert space including deconvolution for continuous signals. 
The theoretical discussion developed in this paper 
naturally leads to a novel practical signal recovery algorithm.
 As a demonstration, a MIMO detection algorithm based on the projected Landweber algorithm is derived.
The proposed MIMO detection algorithm achieves much smaller symbol error rate compared with the MMSE detector.
\end{abstract}

\section{Introduction}

A number of signal detection problems in wireless communications and signal processing 
can be classified into linear inverse problems.
In a linear inverse problem, a source signal $x \in \mathbb{F}^n$ 
($\mathbb{F} = \mathbb{R}$ or $\mathbb{F} = \mathbb{C}$)
is inferred from the noisy linear observation $y = H x + w$ where
$H \in \mathbb{F}^{m \times n}$ and the additive noise $w \in \mathbb{F}^{m}$.

One of the simplest approaches for the above task is to rely on the 
least square principle, i.e., one can minimize $(1/2)\|y - H x \|^2$
which corresponds to the maximum likelihood estimation rule for estimating the source signal $x$
under the Gaussian noise assumption.
The {\em Landweber algorithm}  \cite{Landweber51} \cite{Combettes11} is defined by the fixed-point iteration
\begin{equation} \label{Landweber}
	x^{(k+1)} = x^{(k)} - \omega H^\dag ( H x^{(k)} - y),\ k = 0, 1, 2,\ldots 
\end{equation}
where $H^\dag := H^T$ if $\mathbb{F} = \mathbb{R}$, otherwise $H^\dag := H^H$.
The notation $X^H$ indicates the Hermitian transpose of $X$ and 
the parameter $\omega$ is a real constant. Note that the Landweber algorithm can be 
defined not only on a finite dimensional Euclidean space but also on 
an infinite dimensional complex/real Hilbert space. The Landweber algorithm defined on 
a Hilbert space is especially important for linear inverse problems
including convolutions of continuous signals. 

The Landweber algorithm can be regarded as a gradient descent method for 
minimizing the objective function $(1/2)\|y - H x \|^2$ because $H^\dag ( H x^{(k)} - y)$ 
is the gradient of the objective function.
The Landweber algorithm has been widely employed as a signal reconstruction algorithm 
for image deconvolution \cite{Vankawala86},  inverse problems regarding diffusion 
partial differential equations \cite{Yang17},  MIMO detectors \cite{Zhang17}, and  so on. 
An advantage of the Landweber algorithm is that we can easily include
a proximal or projection operation that utilizes 
the prior knowledge on the source signal after 
the gradient descent step (\ref{Landweber}), which is often called
the {\em projected Landweber algorithm} \cite{Combettes11}.

One evident drawback of the Landweber algorithm is that 
the convergence speed is often too slow and we need to exploit an appropriate 
acceleration method. 
Recently, Takabe and Wadayama \cite{Takabe20}
found that a step-size sequence determined from 
Chebyshev polynomials can accelerate the convergence 
of gradient descent algorithms. Wadayama and Takabe \cite{Wadayama20}
generalized the central idea of \cite{Takabe20} for general fixed-point iterations.
The method is called the {\em Chebyshev inertial iteration}.
It would be natural to apply the Chebyshev inertial iteration for 
improving the Landweber algorithm in terms of the convergence speed.

The successive over relaxation (SOR)
is a common method for accelerating a fixed-point iteration 
$
	x^{(k+1)} = f(x^{(k)}) 	
$
for solving linear equation
such as Jacobi method.
The SOR iteration corresponding to the above equation is given by
$
x^{(k+1)} = x^{(k)} + \omega_k\left( f( x^{(k)} ) - x^{(k)}  \right) 
$
where $\omega_k \in (0, 2)$. 
A fixed SOR factors $\omega_k = \omega$ is commonly used in practice. 
The Chebyshev inertial iteration \cite{Wadayama20} is a natural generalization of 
the SOR method, which is a method employing $\{\omega_k\}$ defined based on the Chebyshev polynomials.
The fundamental properties including the convergence rate 
are analyzed in \cite{Takabe20, Wadayama20}.

The goal of this paper is twofold. 
The first goal is to extend the theory of the Chebyshev inertial iteration
to the Landweber algorithm on Hilbert spaces. 
The arguments in \cite{Takabe20, Wadayama20} are restricted to the case where the underlying 
space is a finite dimensional Euclidean space.
By extending the argument to a Hilbert space, 
the essential idea of the Chebyshev inertial iteration becomes applicable 
to iterative algorithms for infinite dimensional linear inverse problems,
and we can obtain accelerated convergence of these algorithms.
The Chebyshev inertial Landweber algorithm presented in this paper 
can be applied to wide class of signal reconstruction 
algorithms on complex/real Hilbert spaces such as deconvolution for continuous signals.

The second goal is to present a novel MIMO detection algorithm 
based on the projected Landweber algorithm with the Chebyshev inertial iteration for 
demonstrating that the theoretical discussion developed in this paper 
naturally leads to a novel practical signal detection algorithm.

\section{Preliminaries}
In this section, several basic facts on functional analysis required for the
subsequent argument are briefly reviewed. 
Precise definitions regarding functional analysis presented in this section 
can be found in \cite{Muscat14}.
\subsection{Hilbert space}

Let $p$ be a real number satisfying $1 \le p < \infty$.
The set of infinite sequences $(a_1, a_2, \ldots)$ in $\mathbb{F}$ satisfying 
$
	\sum_{i = 1}^\infty |a_i|^p < \infty
$
is denoted by $l^p$ where $\mathbb{F} = \mathbb{C}$ or $\mathbb{R}$.  
If the length of sequences
are finite, i.e., a complex or real vector space, we can define a finite dimensional 
norm space $l^p(N)$.
The set of measurable functions defined on the closed interval  $[a,b] \subset \mathbb{R}$,
$
	\int_{a}^{b} |f(x)|^p dx < \infty
$
is denoted by $L^p(a,b)$. 


Let ${\cal H}$ be a complex or real vector space.
For any $f, g \in {\cal H}$, an inner product $\langle f, g \rangle$
is assumed to be given. 
The norm (inner product norm) associated with the inner product is defined by 
$
	\|f\| := \sqrt{\langle f, g \rangle},\quad f \in {\cal H}.
$
If the norm space $({\cal H}, \|\cdot \|)$ is complete,  
the space is said to be
{\em Hilbert space}.
The norm spaces $l^2$ and $L^2(a,b)$ are Hilbert spaces.
In the case of $l^2$,  the inner products are defined by 
$
	\langle f, g \rangle := \left(\sum_{i=1}^\infty f_i \overline{g_i} \right)^{1/2}.
$
On the other hand, the inner product for $L^2(a,b)$ is defined by
$
	\langle f, g \rangle := \sqrt{\int_{a}^{b} f(x) \overline{g(x)} dx }.
$

Let $T$ be a bounded linear operator on ${\cal H}$. 
The operator norm of $T$ is defined by
$
	\|T\| := \sup_{\|f\| = 1}\|T f\|
$
where $f$ is an element of ${\cal H}$. 
For any $x \in {\cal H}$ and an operator $T$, we have the sub-multiplicative inequality 
\begin{equation} \label{op_ineq}
	\|T x \| \le \|T \| \| x \|.
\end{equation}

\subsection{Compact operator and spectral mapping theorem}

Let $T$ be a linear operator on ${\cal H}$.
A complex number $\lambda \in  \mathbb{C}$
is an eigenvalue of $T$ if and only if 
a nonzero vector $x$ in ${\cal H}$ satisfies 
$
	T x = \lambda x
$
where $x$ is said to be an eigenvector corresponding 
to $\lambda$. In the following,  the set of 
all eigenvalues of $T$ is denoted by $\sigma(T)$.
The spectral radius of $T$ is given as
\begin{equation}
	\rho(T) := \max \{|\lambda | : \lambda \in \sigma(T) \}.
\end{equation}

Suppose that $K(x,y)$ satisfies 
$
	\int_{a}^{b} \int_{a}^{b} |K(x,y)| dx dy < \infty.
$
For any $f \in L^2(a,b)$, let 
$
	T f := \int_{a}^{b} K(x,y) f(y) dy,
$
which is a linear operator on $L^2(a,b)$
and the operator $T$ is a compact operator \cite{Muscat14}. 
If the kernel function $K(x,y)$
satisfies 
 $
	K(x,y) =  \overline{K(y, x)}, 	
$
then the operator $T$ is a {\em compact self-adjoint operator}.
In the case of the finite dimensional space $l^2(N)$, 
a linear operator defined by a Hermitian matrix $T$
corresponds to a compact self-adjoint operator.

The next theorem plays an important role in our analysis described below.
\begin{theorem}[Spectral mapping theorem \cite{Muscat14}] 
Let $T$ be a compact operator defined on ${\cal H}$.
If a complex valued function $f:\mathbb{C} \rightarrow \mathbb{C}$
is analytic around $\rho(T)$, then 
the set of eigenvalues of $f(T)$ is given by
$
	\{f(\lambda): \lambda \in \sigma(T) \}.
$
\end{theorem}

Let $T$ be a compact self-adjoint operator on ${\cal H}$.
It is known that the set of eigenvalues of a compact self-adjoint operator 
is a countable set of real numbers \cite{Muscat14}.
Let $\sigma(T) := \{\lambda_1, \lambda_2, \ldots\} (\lambda_i \in \mathbb{R} )$.
A polynomial $f(x) := a_n x^n + a_{n-1} x^{n-1} + \cdots + a_0$ defined on $\mathbb{C}$
is analytic over whole $\mathbb{C}$.
As a consequence of the spectral mapping theorem, the set of eigenvalues of $f(T)$
is given as 
$
	\sigma(f(T)) = \{f(\lambda_1), f(\lambda_2), \ldots\}.
$
For a compact self-adjoint operator $T$,  we also have 
\begin{equation} \label{norm_spec}
	\|f(T)\| = \rho(f(T)).
\end{equation}


\subsection{Landweber iteration}

Assume that $x$ is an element in a Hilbert space.
The aim of the Landweber iteration is to recover the original signal $x$
from a measurement $y = T x$ or noisy measurement $y$ where $T$ is a linear operator.
The Landweber iteration is a gradient descent algorithm in a Hilbert space to minimize 
the error functional
$
	(1/2) \| T x - y \|^2.
$
The Landweber iteration is 
defined by the fixed-point iteration
\begin{equation}
	x^{(k+1)} = x^{(k)} - \omega T^* ( T x^{(k)} - y).
\end{equation}

\section{Chebyshev Inertial Iteration}

\subsection{Fixed-point iteration}

Suppose that we have a fixed-point iteration defined on a Hilbert space ${\cal H}$:
\begin{equation} \label{org_iter}
	x^{(k+1)} = A x^{(k)} + b,\quad  k = 0, 1, 2, \ldots,
\end{equation}
where $x^{(k)}, b \in {\cal H}$.
The operator $A:{\cal H} \rightarrow {\cal H}$ is a compact self-adjoint 
operator on ${\cal H}$.

Assume also that $A$ is a contraction mapping which
has a fixed point $x^* \in {\cal H}$ satisfying 
$
	x^* = A x^* + b,
$
where the existence of the fixed point is guaranteed by Banach fixed-point theorem \cite{Muscat14}.
The {\em inertial iteration} \cite{Wadayama20} corresponding 
to the original iteration (\ref{org_iter}) is 
defined by
\begin{equation} \label{sor_iter}
	x^{(k+1)} = x^{(k)} + \omega_k \left(A x^{(k)} + b - x^{(k)}  \right),
\end{equation}
where $\{ \omega_k  \}$ are called the {\em inertial factors}.
It should be remarked that the fixed point of the inertial iteration exactly coincides with the
fixed point of the original iteration  (\ref{org_iter}).

\subsection{Analysis on spectral radius}

From the inertial iteration (\ref{sor_iter}), we have the equivalent update equation 
\begin{equation} \label{modified_sor}
	x^{(k+1)} = (I - \omega_k B) x^{(k)} + \omega_k b,
\end{equation}
where $B := I - A$. From the assumption that $A$ is a compact self-adjoint operator, 
we can show $B$ is also compact self-adjoint.
Since the fixed point $x^*$ satisfies 
\begin{equation} \label{fixed}
	x^{*} = (I - \omega_k B) x^{*} + \omega_k b,
\end{equation}
for any nonnegative $k$ and subtracting (\ref{fixed}) from (\ref{modified_sor}), 
we immediately have 
a recursive formula on the residual errors:
\begin{equation} \label{err_recur}
	x^{(k+1)} - x^* = (I - \omega_k B) (x^{(k)} - x^*).
\end{equation}

In the following discussion, we will assume the following inertial factors satisfying 
the periodical condition:
\begin{equation}
	\omega_{\ell T + j} = \omega_j, \quad \ell = 0,1,2,\ldots,\quad j = 0, 1, \ldots, T-1,
\end{equation}
where a positive integer $T$ represents the period.

Recursively applying  (\ref{err_recur}) multiple times, we can get 
\begin{equation}
	x^{(T)} - x^* = \left(\prod_{k=0}^{T-1}(I - \omega_k B)  \right) (x^{(0)} - x^*).
\end{equation}
The norm of the left-hand side can be upper bounded by 
\begin{eqnarray}
	\| x^{(T)} - x^* \| &=& \left\|\left(\prod_{k=0}^{T-1}(I - \omega_k B)  \right) (x^{(0)} - x^*) \right\|  \\
	&\le& \left\|\prod_{k=0}^{T-1}(I - \omega_k B)  \right\| \|  x^{(0)} - x^* \|  \\ \label{req}
	&=& \rho\left(\prod_{k=0}^{T-1}(I - \omega_k B)  \right) \|  x^{(0)} - x^* \|.
\end{eqnarray}
The above inequality is based on the sub-multiplicative inequality (\ref{op_ineq}) and 
the last equality is due to (\ref{norm_spec}).


\subsection{Chebyshev inertial factors}

As we discussed, the operator $B$ is a compact self-adjoint operator and 
it has countable real eigenvalues, 
which are denoted by ${\lambda_k} (k=1,2,\ldots)$. Hereafter, the minimum and the maximum 
eigenvalues of $B$ are denoted by $l_{min}$ and $l_{max}$, respectively.

In the following discussion, we will use the inertial factors
defined by
\begin{equation}
	\omega_k^* := \left[ \lambda_+ + \lambda_{-} \cos \left(\frac{2k+1}{2 T}  \pi \right) \right]^{-1},\  k = 0, 1, \ldots, T-1,
\end{equation}
where
$
	\lambda_{+} := (l_{max} + l_{min})/2,\quad  
	\lambda_{-} := (l_{max} - l_{min})/2.
$
These inertail factors are called the {\em Chebyshev inertial factors}.
A Chebyshev inertial factor is the inverse of a root of a Chebyshev polynomial \cite{Takabe20, Wadayama20}.

Let $\beta_T(x) := \prod_{k=0}^{T-1}(1 - \omega_k^* x)$.
If the Chebyshev inertial factors are employed in the inertial iteration, 
we can use Lemma 2 proved in \cite{Wadayama20} to bound $|\beta_T(\lambda)|$ as
\begin{equation}
|\beta_T(\lambda) | \le \text{sech} \left( T \cosh ^{-1}\left( \frac{\lambda_{+}}{\lambda_{-}} \right) \right).
\end{equation}
for $l_{min} \le \lambda \le l_{max}$.
Combining the spectral mapping theorem 
and the above inequality, we immediately obtain
\begin{equation}
\rho(\beta_T(B) ) = \rho \left(\prod_{k=0}^{T-1} (I - \omega_k B) \right) 
 \le \text{sech} \left( T \cosh ^{-1}\left( \frac{\lambda_{+}}{\lambda_{-}} \right) \right).
\end{equation}
The above argument can be summarized in the following theorem.
\begin{theorem} \label{convergence_rate}
Let $T$ be a positive integer. 
For $\ell = 1,2,3, \ldots$, the Landweber iteration with 
the Chebyshev inertial factors satisfies the residual error bound: 
\begin{equation}
	\|x^{(\ell T)} - x^*\|
	\le  U(T)^\ell \|x^{(0)} - x^*\|,
\end{equation}
where $U(T) := \text{sech} \left( T \cosh ^{-1}\left( {\lambda_{+}}/{\lambda_{-}} \right) \right)$.
\end{theorem}
This theorem implies that the error norm is tightly upper bounded 
if the iteration index is a multiple of $T$.


Combining the Landweber iteration with the Chebyshev inertial iteration,
we have the {\em Chebyshev inertial Landweber algorithm}:
\begin{eqnarray} \nonumber
	x^{(k+1)} &=& x^{(k)} + \omega_k^* \left( x^{(k)} - \omega T^* ( T x^{(k)} - y) - x^{(k)}  \right) \\
	&=& x^{(k)} - \omega_k^* \omega T^*\left( T x^{(k)} - y   \right).
\end{eqnarray}
The iteration is almost the same as the original Landweber iteration except 
for the use of the Chebyshev inertial factors $\{\omega_k^* \}$.
It is common to set the inverse of the maximum eigenvalue of $T^* T$
to the fixed factor $\omega$.

\section{Experiments}

\subsection{Deconvolution by Landweber iteration}

In this subsection, deconvolution by the Landweber iteration
over the Hilbert space ${\cal H} = L^2(-\infty, \infty)$ over $\mathbb{R}$ is studied.
Suppose that $f, g \in {\cal H}$ are  measurable functions on $\mathbb{R}$.
The convolution of $f$ and $g$ are defined as 
\begin{equation} \label{conv_eq}
y(x) := \int_{-\infty}^{\infty} f(u) g(x - u) du,
\end{equation}
which is a compact operator on ${\cal H}$ \cite{Muscat14}. We thus can write 
$
	y = G f
$
where $G: {\cal H} \rightarrow {\cal H}$ is a compact operator defined by (\ref{conv_eq})
and $f, g \in {\cal H}$.
In a context of an inverse problem regarding a linear system involving a convlution, 
the function $f$ can be considered as a source signal and 
$g$ represents a point spread function (PSF) or impulse response. 
In the context of a partial differential equation, $g$ represents 
the Green's function or the integral kernel corresponding to 
a given partial differential equation.

We further assume that $g$ is an even function satisfying $g(x) = g(-x)$.
This implies that the operator $G$ is a compact self-adjoint operator.
In the following, we try to recover the source signal $f$ from the blurred
signal $y$ by using the Landweber iteration on ${\cal H}$ given by
$
	s^{(k+1)} = s^{(k)} - \omega G^* ( G s^{(k)} - y), 	
$
where the initial condition is 
$
s^{(0)} := y.
$
Note that $G^* = G$ holds due to the assumption on $g$.

\begin{figure}[htbp]
\begin{center}
\includegraphics[scale=0.35]{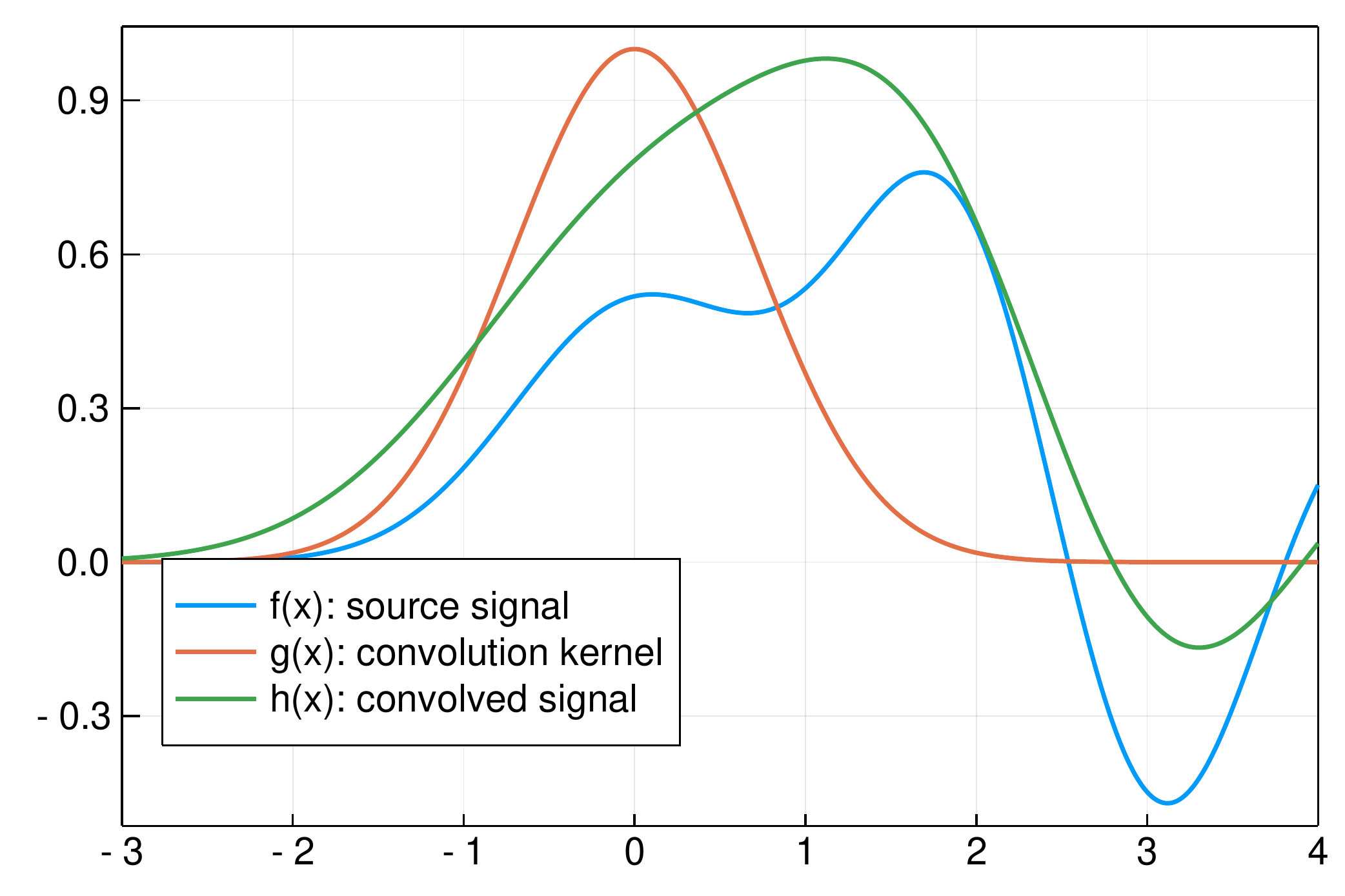}	
\end{center}
\caption{Plots of the source signal $f$, convolutional kernel $g$, and
the convolved signal $y$. The definitions of $f$ and $g$ are given by
$f(x) :=  \frac 1 2 \exp(-x^2) + \exp(-(x-2)^2)
-  \exp(-(x-3)^2)  + \frac 1 2 \exp(-(x-4)^2), 
g(x) := \exp(-x^2).$
}
\label{fig:org_signal}
\end{figure}

In the following experiments shown below, 
the closed range from $-8.192$ to $8.192$ are discretized into 16384 bins,
and convolution integration (\ref{conv_eq}) is approximated by cyclic convolution 
with 16384-points FFT and frequency domain products.
Figure \ref{fig:org_signal} presents the source signal $f$, the convolutional kernel $g$, and
the convolved signal $y$. 

\begin{figure}[htbp]
\begin{center}
\includegraphics[scale=0.35]{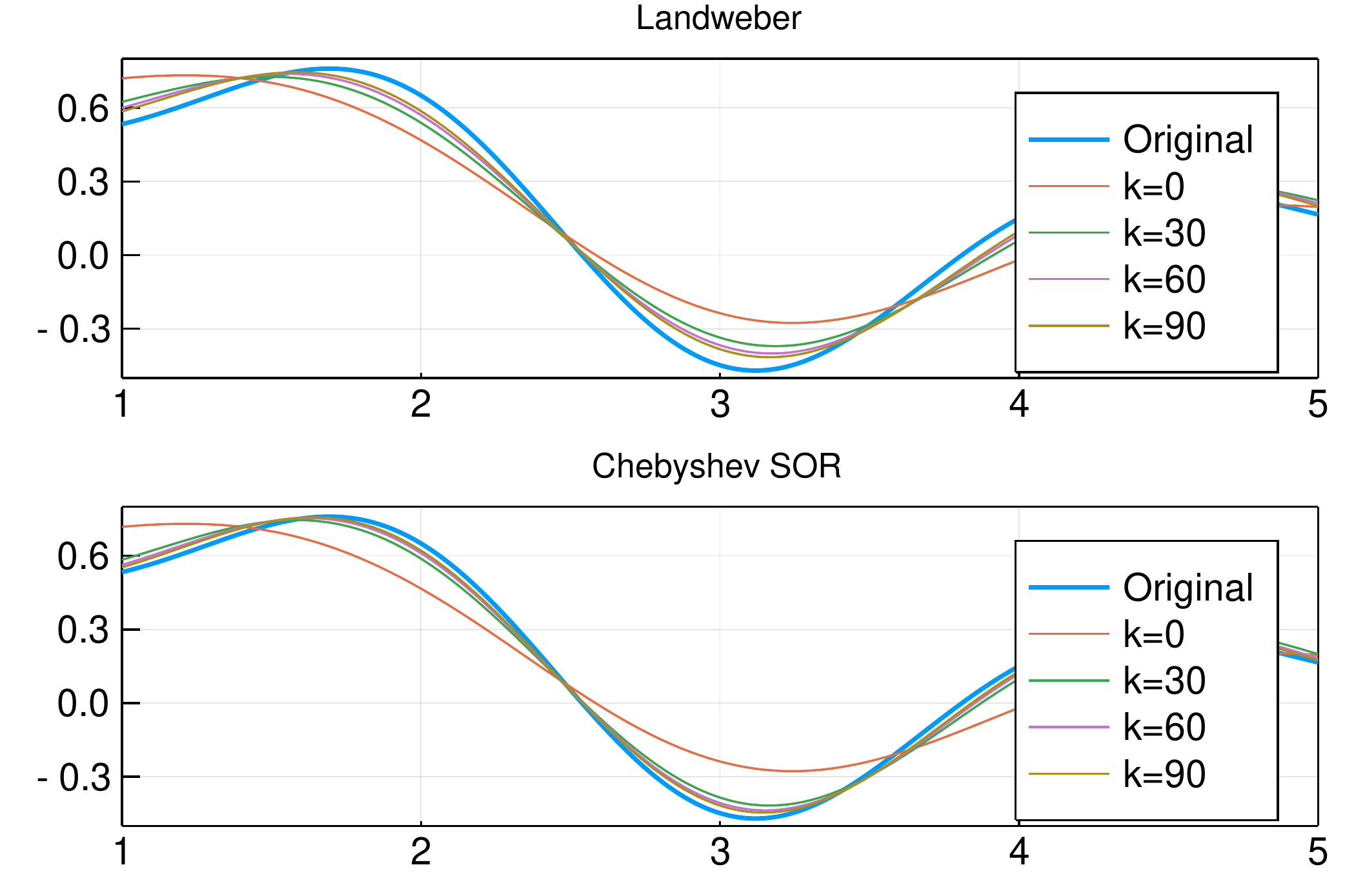}	
\end{center}
\caption{Deconvolution process: (upper) original Landweber iteration, (lower) Chebyshev inertial Landweber iteration. 
The coefficient $\omega$ is set to $0.3$ in the both cases. For Chebyshev inertial Landweber algorithm, 
$l_{min} = 0.1$ and $l_{max} = 0.9$
are used. The period $T$ is set to $8$. The index $k$ represents the number of iteration.}
\label{fig:convergence}
\end{figure}

\begin{figure}[htbp]
\begin{center}
\includegraphics[scale=0.35]{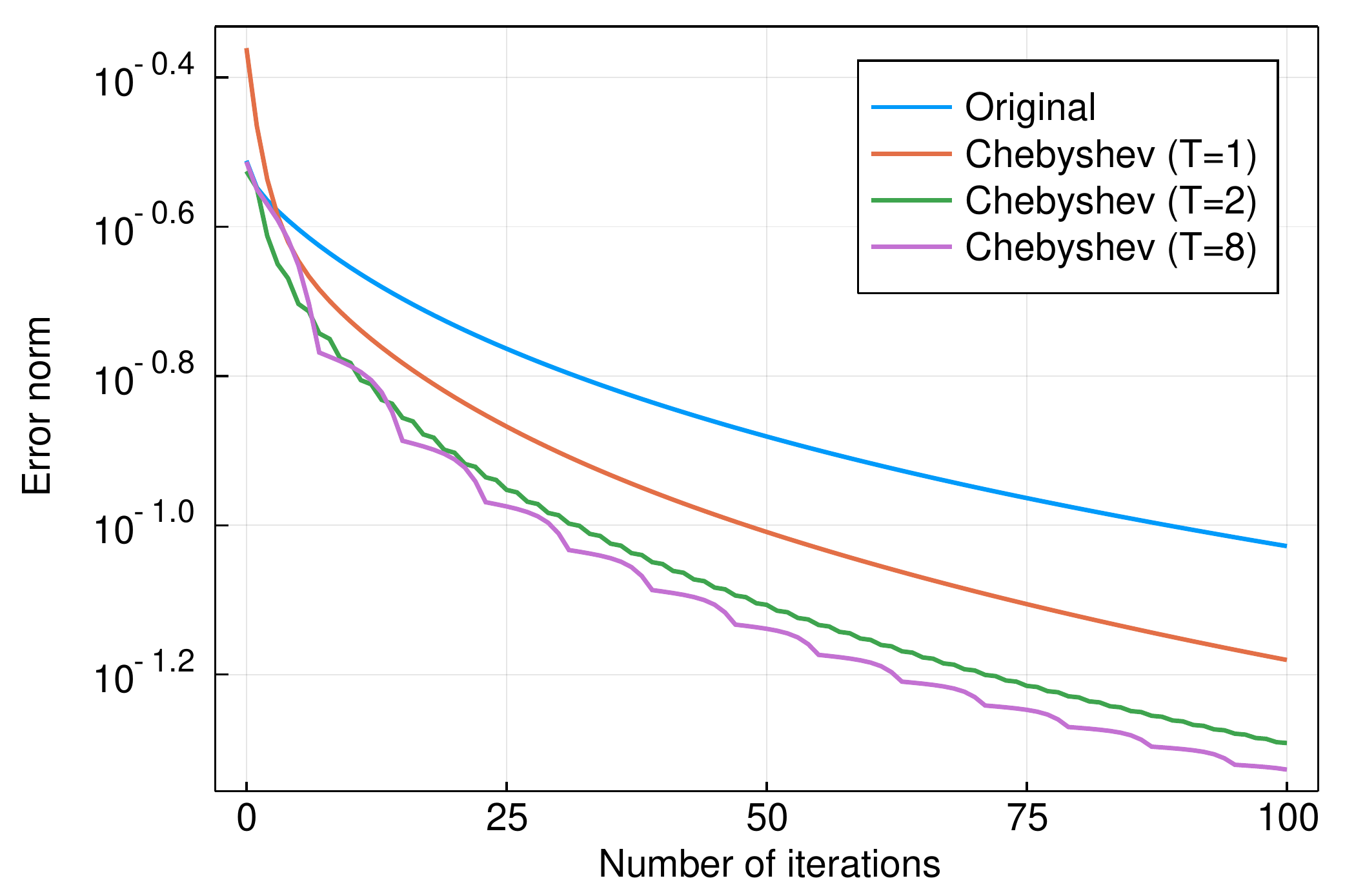}	
\end{center}
\caption{Error norms $||s^{(k)} - f||$ as functions of number of iterations in deconvolution processes.}
\label{fig:err_curve}
\end{figure}

Figure \ref{fig:convergence} presents a deconvolution process 
by the original Landweber algorithm (upper) and the Chebyshev inertial Landweber algorithm (lower).
The tentative results $s^{(k)}$ for every 30 iterations are shown.
In both cases, we can observe that $s^{(k)}$ gradually approaches to the 
source signal $f$.  This is because $I - \omega G^* G$ is a contractive mapping under the setting 
$\omega = 0.3$. Comparing two figures, we can see that the Chebyshev inertial Landweber algorithm
shows much faster convergence to the source signal $f$.
This observation is also supported by the error curves depicted in Fig.\ref{fig:err_curve}.
The Chebyshev inertial Landweber algorithm $(T= 1, 2, 8)$ shows 
faster convergence 
compared with the original Landweber iteration. 
For example, the error $||s^{(k)} - f|| = 0.1$ is achieved with 
$k = 100$ with the original Landweber iteration but the Chebyshev iteration 
requires only $k = 25$ iterations.

\subsection{Finite-dimensional least square problem}

In this subsection, we will discuss a least square problem closely related to 
MIMO detection problems. We here assume a finite dimensional Hilbert 
space ${\cal H} = l^2(n)$ on $\mathbb{C}$.

Let $H \in \mathbb{C}^{n \times n}$ be a complex matrix whose 
elements follow the normal complex Gaussian distribution ${\cal CN}(0, 1)$.
Our task is to recover a source signal $x \in \mathbb{C}^n$ from 
a noisy linear measurement $y = H x + w$ where
$w$ is a complex Gaussian noise vector. 
The problem can be seen as a MIMO detection problem where the numbers of transmit and
receive antennas are $n$.
The least square problem 
$
	\hat x := \text{minimize}_{x \in \mathbb{C}^n} (1/2) \| y - H x \|^2
$
is equivalent to the maximum likelihood estimation rule for the above setting.

In this case, the Landweber iteration is exactly same as a gradient descent process
\begin{equation}\label{finite_landweber}
	s^{(k+1)} = s^{(k)} - \omega H^H ( H s^{(k)} - y)		
\end{equation}
for minimizing the objective function.
It is known that the asymptotic convergence rate is optimal if
$
 	\omega = \omega_{opt} := {2}({\lambda_{min}(H^H H) + \lambda_{max}(H^H H)})^{-1}	
$
holds. In this case, the matrix 
$
	A := I - \omega_{opt} H^H H
$
becomes a Hermitian matrix and thus $A$ is a compact self-adjoint operator on ${\cal H}$.
This means that we can apply the Chebyshev inertial iteration to 
the Landweber iteration.

The details of the experiment are summarized as follows.
The dimension $n$ is set to $32$. 
Each element of $x$ is chosen from 8-PSK constellation 
$\{ \exp((2\pi j k)/8) \} (k = 0,1,\ldots, 7)$ uniformly at random. 
The optimal factor  $\omega_{opt}$ is employed in the Landweber iteration.
Each element of the noise vector $w$ follows ${\cal CN}(0, \sigma^2)$ 
where $\sigma = 10^{-4}$.
The minimum and maximum eigenvalues of $\omega_{opt} H^H H$ are 
used as $l_{min}$ and $l_{max}$ for determining the Chebyshev inertial factors.

\begin{figure}[htbp]
\begin{center}
\includegraphics[scale=0.35]{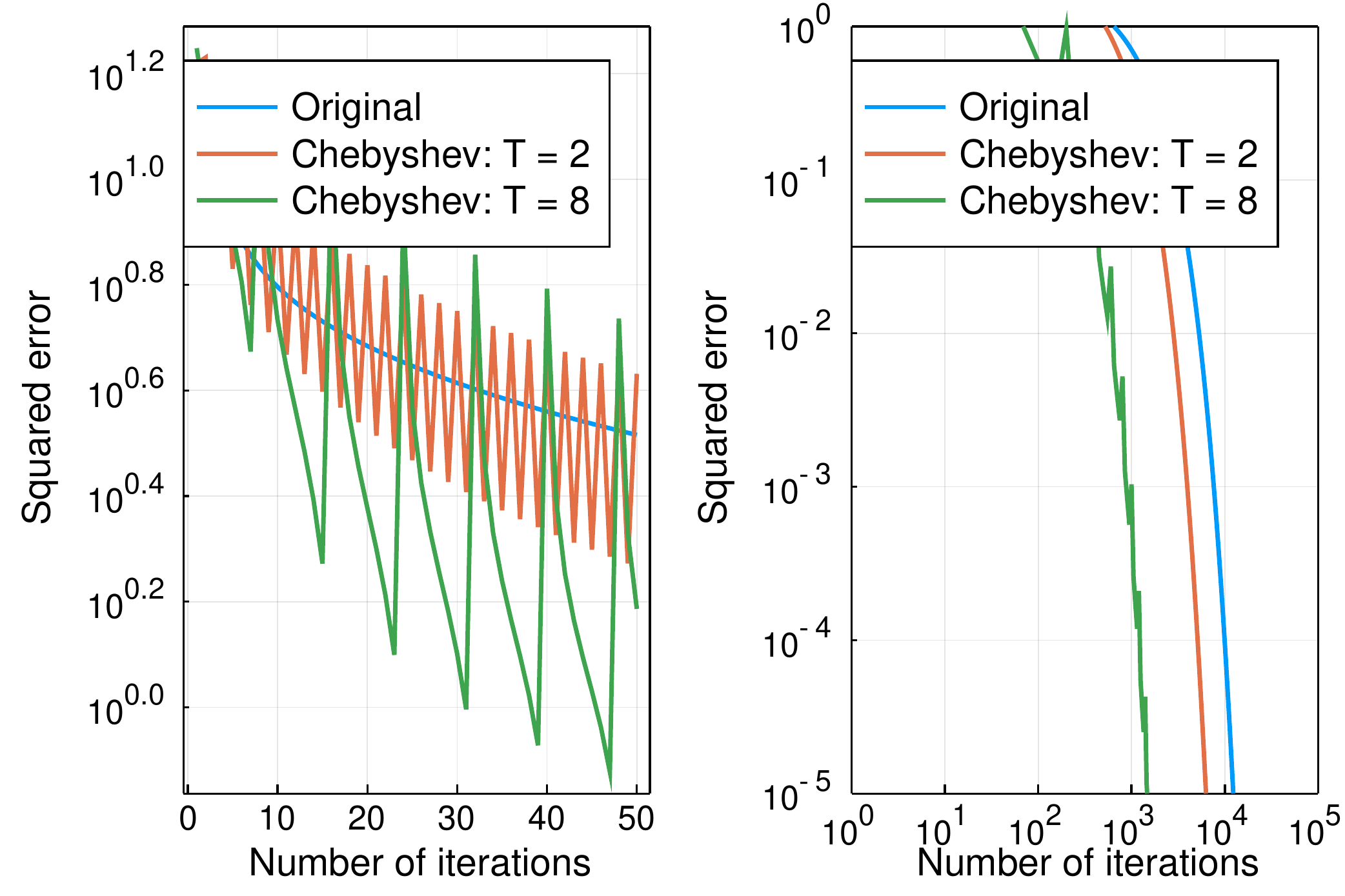}
\end{center}
\caption{Squared error norms $\|s^{(k)} - x \|^2$  as a function of number of iterations: (left) number of iteration is up to 50, (right) number of iteration is up to 30000.
The squared error norms are averaged for 100-trials. }
\label{fig:MIMO_Landweberl}
\end{figure}

Figure \ref{fig:MIMO_Landweberl} summarizes the comparison on 
the squared error norms $\|s^{(k)} - x \|^2$
between the original Landweber iteration and the accelerated iterations.
From Fig. \ref{fig:MIMO_Landweberl} (left), 
we can immediately recognize the zigzag-shaped error curves of the Chebyshev inertial Landweber algorithm.
This is because the error norm is exponentially upper bounded only 
when the iteration index $k$ is multiple of $T$ as shown in Theorem \ref{convergence_rate}. 
It mean that the lower envelope of the zigzag curves can be seen 
as the guaranteed performance of the Chebyshev inertial Landweber algorithm.
Figure \ref{fig:MIMO_Landweberl} (right) indicates the squared errors up to $k=30000$.
The proposed schemes (Chebyshev $T = 2, 8$) shows much faster convergence compared with the original 
Landweber iteration.

\begin{figure}[htbp]
\begin{center}
\includegraphics[scale=0.35]{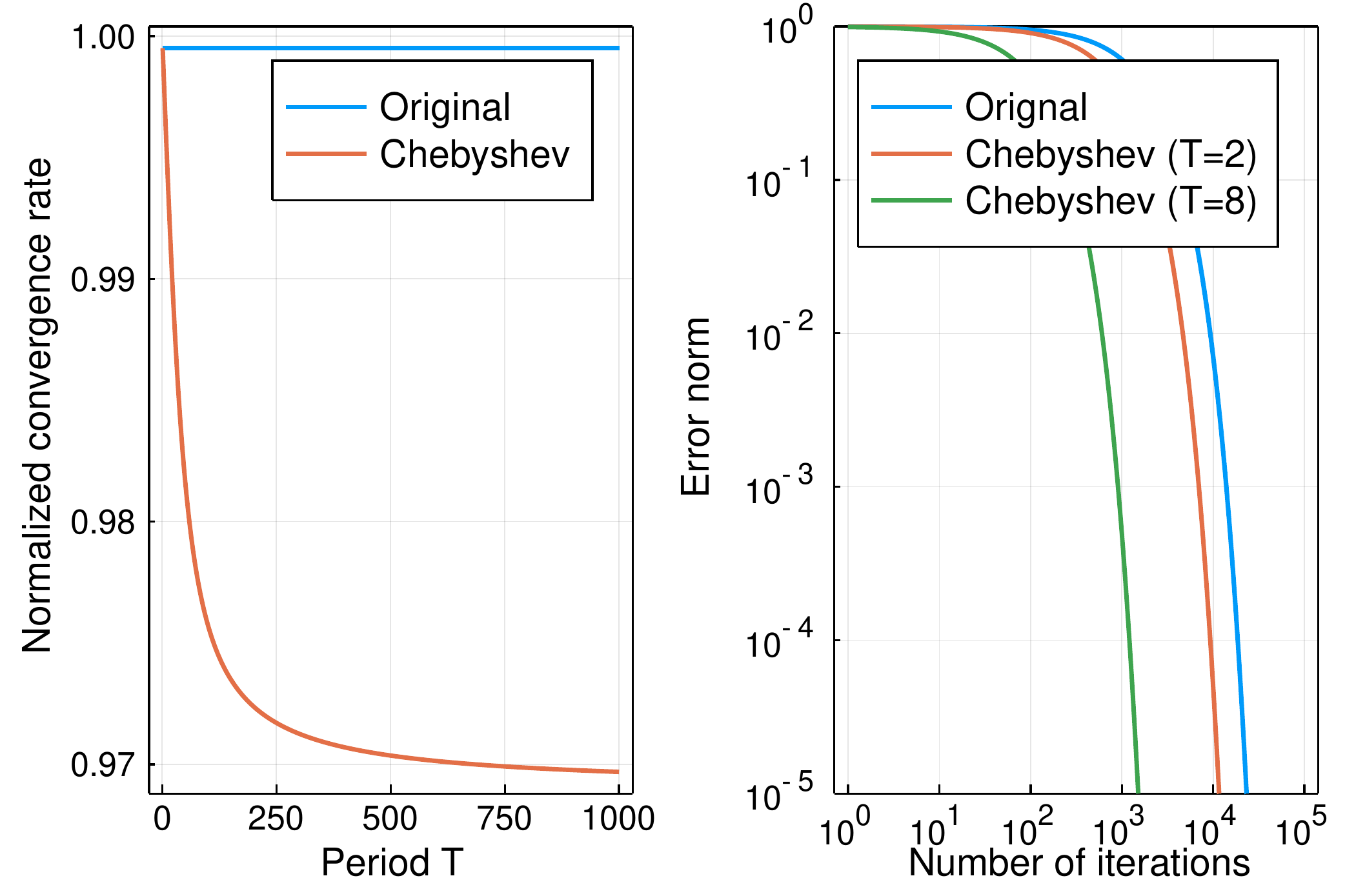}
\end{center}
\caption{(left) Normalized convergence rate $\rho(A)$ and $U(T)$, (right) approximate error norms derived from $\rho(A)$ and $U(T)$: the approximated error norms are 
$\rho(A)^k, U(2)^k, U(8)^k$, respectively. }
\label{fig:MIMO_Theory}
\end{figure}

The speed of convergence of the Landweber iteration is dominated by the 
spectral radius 
$
	\rho(A) = \rho \left(I - \omega_{opt} H^H H \right).
$
On the other hand, $U(T)$ indicates an upper bound of the normalized 
asymptotic convergence rate depending on the period $T$.
Figure \ref{fig:MIMO_Theory} (left) presents both $\rho(A)$ and $U(T)$ as functions of $T$.
As $T$ becomes larger, the value of $U(T)$ becomes strictly smaller than $\rho(A)$. 
This means that the Chebyshev inertial iteration certainly accelerates the asymptotic convergence rate.
Figure \ref{fig:MIMO_Theory} (right) shows
approximate error norms derived from $\rho(A)$ and $U(T)$ which are given by
$\rho(A)^k, U(2)^k, U(8)^k$. Although these values does not include the initial errors $\|s_0 - x\|$,
these curves qualitatively explains the behavior of the actual error curves 
in Fig. \ref{fig:MIMO_Landweberl} (right). These results support the usefulness of 
the theoretical analysis discussed in the previous section.

\subsection{Projected Landweber-based MIMO detection}

It is natural to consider the projected Landweber algorithm \cite{Combettes11} for 
making use of the prior information of the source signal to 
improve the reconstruction performance.
We here examine the detection performance of 
a simple projected Landweber-based MIMO detection algorithm.

\begin{figure}[htbp]
\begin{center}
\includegraphics[scale=0.35]{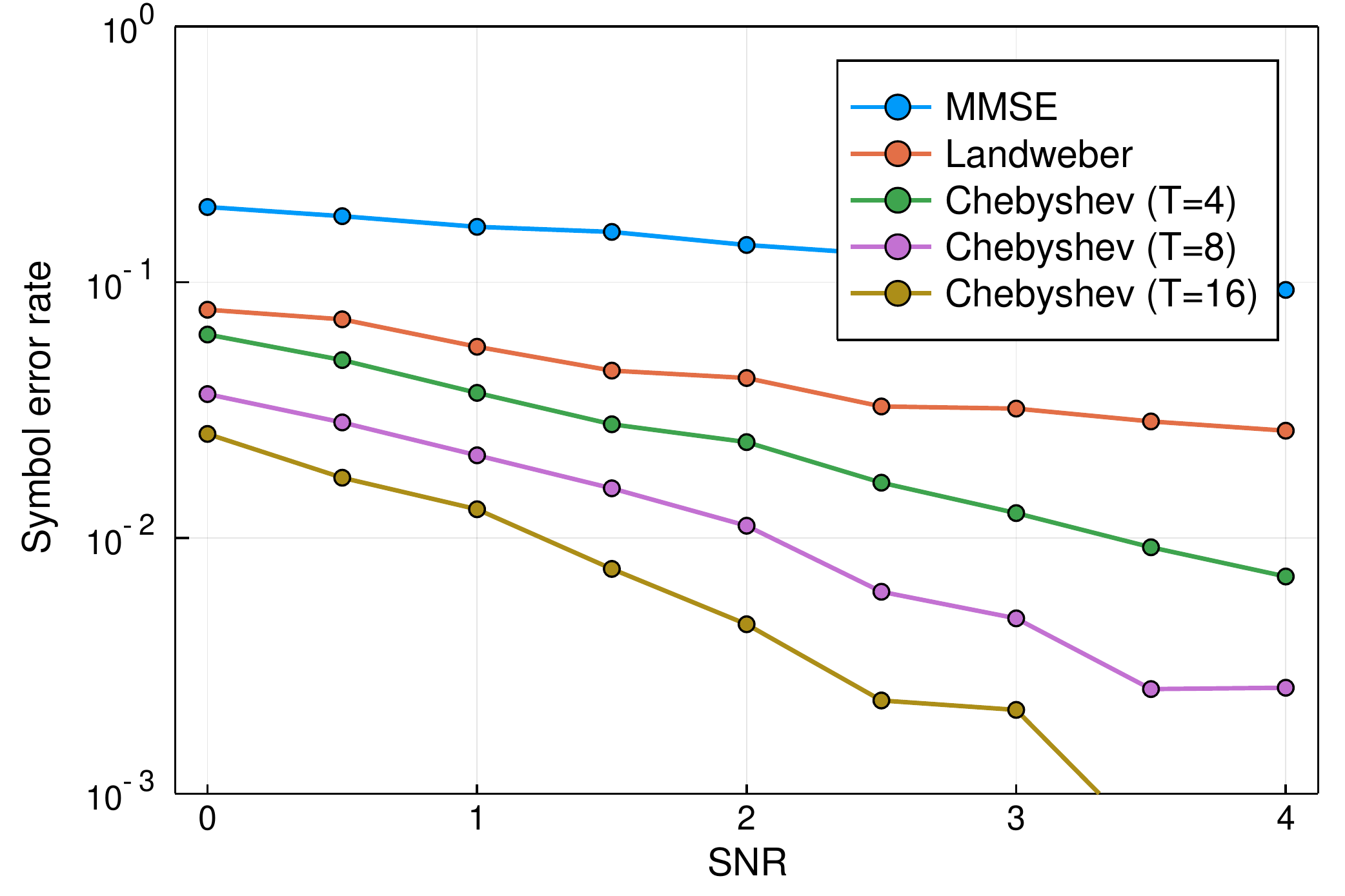}
\end{center}
\caption{Symbol error rate of the projected Landweber algorithm with the Chebyshev inertial iteration. 
The channel model is $32 \times 32$  MIMO channel with 8-PSK input. 
The number of iterations for all the algorithms (except for MMSE) is set to 100.
As baselines,  the performance of the MMSE detector defined by
$
	\hat{x} := (H^H H + \sigma^2 I)^{-1} H^H y
$ is included. The SNR $snr$ (in dB) and $\sigma$ is related as $\sigma = \sqrt{10^{-snr/10}}$.
In the projected Landweber algorithms, $\alpha^2 = 0.5$ is used when iteration index $k$ is less than 20;
for $k \ge 20$, $\alpha^2 = 0.25$ is used.
 }
\label{fig:MIMO_SER}
\end{figure}

The soft projection operator $\eta: \mathbb{C} \rightarrow \mathbb{C}$ used here 
is defined by 
\begin{equation}
	\eta(r) := \frac{\sum_{p \in S} p \exp\left({-| r - p |^2}/{\alpha^2}  \right)}
	{\sum_{p \in S} \exp\left({-| r - p |^2}/{\alpha^2}  \right)},
\end{equation}
where $S \subset \mathbb{C}$ is a signal constellation \cite{Takabe19}.
The iteration of the projected Landweber algorithm is fairly simple;
the soft projection operator is just element-wisely applied to 
the output $s^{(k+1)}$ in (\ref{finite_landweber}) and then, 
the output is passed to the inertial iteration process.
The details of the experiments are as follows.
The channel model is exactly the same as that used in the previous subsection 
($32 \times 32$ MIMO channel with 8-PSK input).
The minimum and maximum eigenvalues $l_{min}$ and $l_{max}$ are determined 
according to the Marchenko-Pastur law.
Figure \ref{fig:MIMO_SER} presents the symbol error rate (SER) of the proposed scheme.
The projected Landweber detectors 
achieves one or two order of magnitude smaller SER than those of the MMSE detector.
The accelerated schemes (Chebyshev $T=4, 8, 16$) provide steeper error curves 
than that of the projected Landweber detector without Chebyshev inertial iteration (denoted by ``Landweber'').
This observation supports the potential of the Chebyshev inertial iteration.

\section*{Acknowledgement}
This work was supported by JSPS Grant-in-Aid for Scientific Research (B) 
Grant Number 19H02138.

\end{document}